\documentclass[]{interact}

\usepackage{epstopdf}

\usepackage[natbibapa,nodoi]{apacite}
\theoremstyle{plain}

\theoremstyle{definition}

\theoremstyle{remark}

\usepackage{tabu}                      
\usepackage{booktabs}         
\usepackage{enumitem}
\usepackage{subcaption}
\usepackage{wrapfig}
\usepackage{colortbl}
\usepackage{xcolor}
\usepackage{wrapfig}
\usepackage{color}
\usepackage{caption}
\usepackage{array}
\usepackage{multicol}
\usepackage{multirow}
\usepackage{tabularx}
\usepackage{todonotes}
\usepackage{xurl}
\usepackage{algorithmic}
\usepackage{algorithm2e}
\usepackage{balance}
\usepackage{lipsum}
\usepackage[indentfirst=false,rightmargin=0pt]{quoting}
\newcommand\cparagraph[1]{\vspace{1.0mm}\noindent\emph{#1}}
\newcommand\bparagraph[1]{\vspace{1.0mm}\noindent\textbf{#1}}
\usepackage{pifont}

\begin{document}


\title{The Dark Side of Augmented Reality: Exploring Manipulative Designs in AR}

\author{
\name{Xian Wang\textsuperscript{a,b}, Lik-Hang Lee\textsuperscript{c}, Carlos Bermejo Fernandez\textsuperscript{a}, and Pan Hui\textsuperscript{a,b,d}\thanks{CONTACT PERSON: Pan Hui and E-MAIL: panhui@ust.hk}}
\affil{\textsuperscript{a}The Hong Kong University of Science and Technology, Hong Kong SAR;  \textsuperscript{b}The Hong Kong University of Science and Technology (Guangzhou), Guangzhou, China;
\textsuperscript{c}The Hong Kong Polytechnic University, Hong Kong SAR; \textsuperscript{d}University of Helsinki, Helsinki, Finland}
}

\maketitle

\begin{abstract}
Augmented Reality (AR) applications are becoming more mainstream, with successful examples in the mobile environment like Pokemon GO. Current malicious techniques can exploit these environments' immersive and mixed nature (physical-virtual) to trick users into providing more personal information, i.e., dark patterns. Dark patterns are deceiving techniques (e.g., interface tricks) designed to influence individuals' behavioural decisions. However, there are few studies regarding dark patterns' potential issues in AR environments. In this work, using scenario construction to build our prototypes, we investigate the potential future approaches that dark patterns can have. We use VR mockups in our user study to analyze the effects of dark patterns in AR. Our study indicates that dark patterns are effective in immersive scenarios, and the use of novel techniques, such as `haptic grabbing' to draw participants' attention, can influence their movements. Finally, we discuss the impact of such malicious techniques and what techniques can mitigate them.
\end{abstract}


\section{Introduction}~\label{sec:introduction}
Augmented Reality (AR) applications seamlessly merge the virtual world into the physical world. We can also see the popularity of AR applications in smartphone applications such as Pokemon Go and Snapchat, as well as other major players in the consumer market such as AR (ARKit), Google (AR Core), and Microsoft (HoloLens)~\citep{patil2019accelerating,ar_market}. 
Moreover, the ubiquity and immersion of these applications open a new space for interaction techniques and the use of these AR applications. For example, the users are not attached to a computer and can freely move in the physical world while interacting with the virtual world or augmented view~\citep{nijholt2021experiencing}. These environments raise new challenges regarding the privacy and security of users~\citep{roesner2014security,hu2021lenscap,guzman2021unravelling}.

AR technologies have the ability to integrate virtual objects with the physical world in order to change the users' perceptions of reality~\citep{chatzopoulos2017mobile}. While these applications are becoming more widespread in the commercial market, there are still limited design guidelines for the AR space. AR application design faces several changes: (i) the rapidly evolving best practices; (ii) challenges with scoping guidelines; (iii) learning by doing~\citep{drascic1996perceptual,azuma1997survey,milgram1994taxonomy,speicher2018xd}. The still newly developed AR technologies (e.g., mid-air interaction, tracking capabilities) can hinder the creation of universal design guidelines. According to Michael Nebeling~\citep{michigan_ux_xr}, we can highlight four different sources that attempt to come up with such design guidelines: (i) vendors, platform-driven; (ii) designers, user-oriented; (iii) practitioners, experience-based; (iv) researchers, empirically derived. The HCI community has been studying the effects of UX in society~\citep{ardito2014investigating,mannonen2014uncovering,rajeshkumar2013taxonomies} and from the perspective of designers and practitioners~\citep{tromp2011design,larusdottir2012big} in order to provide more responsible guidelines and mechanisms for designing UX. Moreover, the current HCI studies~\citep{machuletz2020multiple,mhaidli2021identifying,di2020ui,mathur2019dark,utz2019informed} focus on the influence of such design interfaces on user experiences (UX) and how particular design `tricks' (dark patterns~\citep{darkPatternsBrignull}) can influence the users' decisions towards behaviors desired by the designers. Following the work~\citep{gray2018dark}, we define dark patterns as `\textit{user interface designs that benefit the system instead of the user}'. 

While previous research has evaluated AR security and privacy~\citep{hu2021lenscap,mcpherson2015no,roesner2014security}, the impact of user interfaces on user behavior has received little attention. 
This work aims to propose a first overview of the possible dark pattern designs that the AR ecosystem can have. It is vital 
to address the risks that users might face in AR applications before AR becomes established in mainstream technologies.

In the case of AR scenarios, the possibilities for dark pattern designs are even broader. They 
can put the users at physical risk due to the nature of displaying virtual objects in the physical world~\citep{jung2018ensuring,azuma1997survey}. Many of the current AR applications are still using mobile OS and smartphones as interaction devices~\citep{chatzopoulos2017mobile}, which can also enable traditional dark pattern designs used in other non-AR mobile applications, such as cumbersome menus to disable default settings or misleading information in the 2D menu interfaces~\citep{gray2020kind,gray2018dark,di2020ui,bosch2016tales}. 

\textit{Use case:} ``We can observe dark patterns when the application hides important content, such as menu preferences, behind the user or other physical objects, in order to keep the default values.'' In this work, we 
focus on interface designs that require visualization. However, more importantly, dark patterns in AR applications could be open to other interaction approaches such as haptic feedback that is rendered to nudge 
users to select an option and audio feedback that interrupts or distracts the user during their decision process.

\bparagraph{Contributions.} In this paper, we continue the research on dark patterns by exploring the possibilities of such techniques in AR environments. First, we use the scenario construction~\citep{mhaidli2021identifying} method to illustrate the possible techniques, usages, and risks that dark patterns can enable in the AR ecosystem. 
This allows us to design and develop our dark pattern prototypes in new technologies, such as AR, which can be seen in other works~\citep{mathur2021makes,betten2018constructing} to predict the impact of future technologies. We analyze and describe several novel approaches to applying dark patterns in AR scenarios using the above method. 
Second, following the scenario constructions and building on top of prior works~\citep{mhaidli2021identifying,greenberg2014dark,tseng2022dark}. We use VR as a mock-up environment~\citep{prange2021priview}, where we designed an application that emulates an AR scenario in a public environment (e.g., streets). 
We use the VR mock-up for the following reasons:
\label{sec:reasons}
\begin{enumerate}
\item \textbf{AR technology limitations.} Currently, AR technology is still at its nascent stage. As such, achieving the effect of our prototype requires a significant financial investment and considerable time~\citep{rymarczyk2020technologies}. In contrast, the proposed VR environment allows us to overcome technical issues, for instance, the anchors' tracking and positioning accuracy issue -- virtual objects do not keep the same position)~\citep{yigitbas2021using}. 
\item \textbf{Examples of previous studies.} Moreover, a prior work~\citep{prange2021priview} also leverages a VR mock-up approach to evaluate their AR scenarios. 
\item \textbf{Security and policy.} In our setup, we designed a street view virtual environment that simulates real streets where participants testing in the real world may encounter safety issues (e.g., crossing streets without awareness due to the AR application being tested) not only to the subject but also the bystanders, i.e., we cannot disturb the bystanders causing hassle to the urban environments. It is also important to note that putting a manipulative design in AR would violate university IRB regulations and put participants at risk.
\item \textbf{Control environment variables.} The VR mock-up scenarios allow us to have a consistent setup (e.g., lighting conditions) across all participants. This mock-up can potentially serve as digital twins~\citep{barricelli2022digital} of the reality, that is, a safe and immersive environment for evaluating the proposed manipulative designs.
\end{enumerate}
These findings highlight the importance of guidelines and better mechanisms in AR environments where dark patterns can significantly impact users' behavior. 

\section{Related Work}~\label{sec:related_work}
We organize the related work into two parts according to: AR UX design, where we describe guidelines, frameworks, and techniques in the AR ecosystem; and dark patterns, where we analyze the different taxonomies and how these techniques are used in the digital world.

\subsection{AR UX design ethics}
\label{sec:devices}
AR encompasses the technologies that attempt to augment a physical environment using virtual graphics (e.g., objects, interfaces). The most common approach to displaying this augmented view is using smartphones and smart glasses or headsets~\citep{roesner2014security}. AR applications should be context-aware and adapt to the environment in which they are being used~\citep{ganapathy2013design}. For example, mobile AR applications should have the right form factor for mobile environments~\citep{ganapathy2013design}. In this work, we focus on head-mounted AR devices such as AR glasses and headsets, there are currently two forms of mainstream devices, 1) the images, generated by the device, overlap with the real world through optical see-through displays (e.g. Microsoft Hololens 2\footnote{\url{https://www.microsoft.com/en-us/hololens}}, Magic Leap\footnote{\url{https://www.magicleap.com/en-us/}}); 2) the surroundings are captured through the front camera of the VR device and then the composite image is displayed in the screen  to display the augmented view seamlessly, i.e., video see-through display (e.g. Meta Quest 2\footnote{\url{https://www.meta.com/quest/products/quest-2/}}, Pico 4\footnote{\url{https://www.picoxr.com/global/products/pico4}}). 

AR technologies pose an important risk to users regarding their privacy. The continuous monitoring of the camera sensor can capture information about the users' surroundings, such as their room~\citep{roesner2014security}. Therefore, it is crucial to think about the ethical aspects of UX designs. Dark patterns and design techniques that manipulate users can pose potential risks to their privacy~\citep{gray2018dark,roesner2014security}. The ethical designs of new UI paradigms, such as AR interfaces, are of important focus in the Human-Computer Interaction (HCI) community~\citep{grimpe2014towards}. Mhaidli et al.~\citep{mhaidli2021identifying} analyze the possibilities of manipulative advertisement in XR (eXtended Reality: AR, VR, MR)~\citep{stanney2021extended}. They showcase several scenarios where advertisements target users with a misleading experience, hence influencing their decisions (e.g., by creating artificial emotions in consumers). 
Our work highlights the current problems that such AR-based interfaces can have and evaluate the impact of dark patterns in these immersive scenarios.

\begin{figure}[t]
  \centering
      \includegraphics[width=0.5\columnwidth]{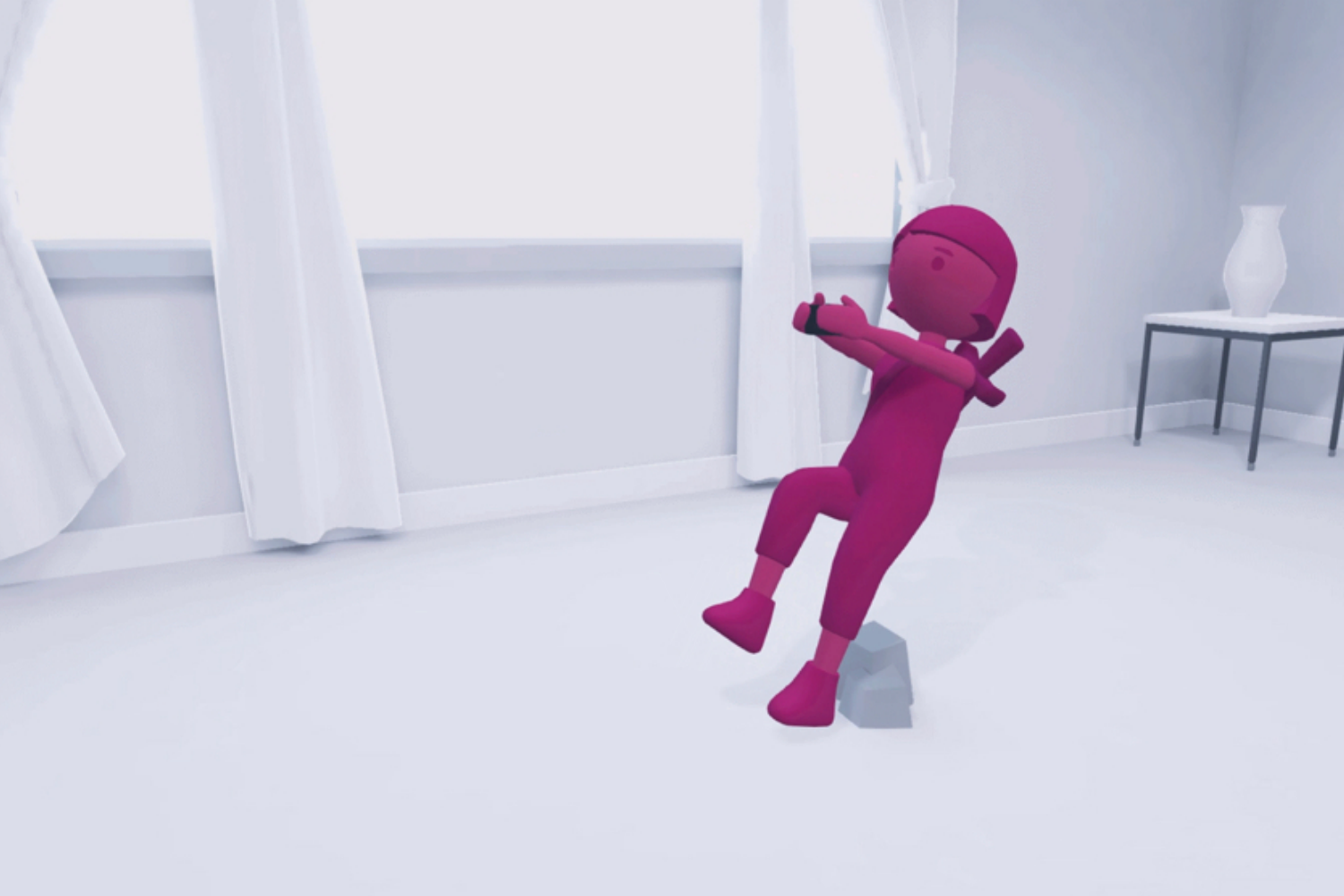}
      \caption{A screenshot captured from the Google ARCore design guidelines. We can see a person tripping over a physical (real) object while walking backward and using an AR application. }
      \label{fig:related_work:googleAR_safety} 
\end{figure}

Moreover, several scholars have highlighted the risks that AR technologies can have on consumers, such as nausea, motion sickness, and safety~\citep{roesner2014security,jung2018ensuring,pierdicca2020augmented}. For example, AR applications can make users fall down when walking towards a virtual object, as they are not focusing on their surroundings~\citep{jung2018ensuring}. As we can observe in several platforms, such as Google ARCore\footnote{\url{https://developers.google.com/ar/design}} and Apple ARKit\footnote{\url{https://developer.apple.com/design/human-interface-guidelines/ios/system-capabilities/augmented-reality/}}, they consider such risks in their design guidelines and other ethical considerations. For example, virtual objects should have an appropriate size that does not force the users to walk backward to see them in their totality, see Figure~\ref{fig:related_work:googleAR_safety}.

Several works emphasize the importance of UX design in order to provide ethical applications~\citep{gray2018dark,ethical_design,ethical_design_karr}. Although any design can have the effect of influencing the user~\citep{oinas2009persuasive,rajeshkumar2013taxonomies}, the abusive use of `malicious' techniques, such as dark patterns, should be restricted~\citep{gray2018dark}. Dark patterns are an example of abusive interface designs that influence the users' welfare and their behavior (e.g., artificially creating a sense of scarcity of flight tickets on online websites).

\bparagraph{Haptics for XR.} Haptic stimuli has been a common approach in mobile devices (smartphones, wearables) to provide a notification channel for mobile applications~\citep{schaub2015design,mehta2016privacy,luk2006role}. We can see haptic feedback broadly used in games, where the controllers provide haptic sensations when the virtual player crashes with another virtual object (e.g., car crash in the game)~\citep{tamborini2006role}. Haptics is also a common technique in other XR scenarios, such as VR games~\citep{bermejo2021exploring}. Haptic devices will play an important role in AR experiences, where the haptic stimuli can enhance the realism and immersion of AR~\citep{bermejo2021survey}. Moreover, haptic stimuli can evoke different emotions according to the feedback displayed (e.g., frequency and intensity of the haptic feedback)~\citep{rovers2004him,obrist2015emotions,vekemans2021motus,eid2015affective}. Users can associate particular valence and arousal according to the received haptic stimuli~\citep{yoo2015emotional,vekemans2021motus}. In~\citep{yoo2015emotional}, the authors showed that the frequency of tactile icons affects people's perception of emotions. Therefore, haptic feedback and the emotions it evokes can be used in the design of dark patterns for XR applications. We focus on certain dark pattern designs in AR with haptic feedback, as it is a commonly used channel to interact with users and can be implemented in current wearables. 

Similarly, the audio sensory channels can be used to enhance the realism and immersion of AR experiences~\citep{chatzopoulos2017mobile}. Although these audio techniques can be considered in the design of dark patterns, we believe that audio should be tested in a physical environment (e.g., streets) and not VR mockups, so noises from the outside world can reduce the performance of audio interaction in AR. In this work, we focus on haptic and visual approaches and consider audio cues as part of future work.

\subsection{Dark patterns}
Dark patterns can be found in highly diversified 
environments, such as online media, games, and social networks~\citep{gunawan2021comparative,fitton2019creating,utz2019informed,mathur2019dark,machuletz2020multiple}. The ubiquitous 
presence of dark patterns becomes even direr when we think about AR environments. As we will see in this paper, the mixed relation between physical (real) and virtual objects in AR applications enables new approaches to `tricking' users. Thus, addressing such threats becomes relevant. 

Brignull~\citep{darkPatternsBrignull} gathers on his website a collection of real-world examples of dark patterns commonly used in websites and mobile applications. The website aims to raise users' awareness of dark pattern practices and constructs a reported dataset of widely used dark patterns. Brignull proposes an initial taxonomy of 12 elements according to the user interface `trick', such as \textit{privacy Zuckering} that tricks users into sharing more personal information. 

Several works propose taxonomies to categorize various 
dark pattern techniques~\citep{conti2010malicious,gray2018dark,mathur2021makes} in a broader sense. Conti~\textit{et al.}~\citep{conti2010malicious} categorize the dark pattern techniques as malicious interface design approaches. They categorize 11 interface designs that can be problematic for users. For example, obfuscation hides interface elements from users. Gray~\textit{et al.}~\citep{gray2018dark} propose a taxonomy using a similar data collection for their corpus as Brignull (tweets from users). They introduced new dark patterns on top of the categories of Brignull, such as nagging and forced action. The latter requires users to perform a specific action repeatedly in order to continue. The authors increase the number of categories in a later version~\citep{gray2020kind} by collecting users' posts about dark patterns on Reddit. Outside the interface design techniques, we can mention the work by~\citep{mathur2021makes} that proposes the addition of normative perspectives in the analysis and categorization of dark patterns. These normative perspectives focus their taxonomy on disciplines outside the HCI community, such as psychology, economics, and ethics. 

The prior works on dark patterns mainly focus on the UIs of traditional systems (e.g., websites, mobile apps), which can be restricted by the aforementioned taxonomies. Studies, e.g., \citep{mathur2021makes,greenberg2014dark} show 
new perspectives that are not limited to 
user interfaces and can be extended to 
new environments, such as AR. For example, the use of haptics, as a dark pattern technique, is not usually included in prior taxonomies. Additionally, 
only some works~\citep{mhaidli2021identifying,greenberg2014dark} illustrate more techniques to manipulate users' decisions.

Our work serves as the groundwork for the design of dark patterns in the emerging AR scenarios, especially in the following two taxonomies from these works: Gray~\textit{et al.}~\citep{gray2018dark}, Mathur~\textit{et al.}~\citep{mathur2021makes}, and Greenberg~\textit{et al.}~\citep{greenberg2014dark}. We leverage 
these taxonomies as they are the most recent. They further highlight 
UI perspectives~\citep{gray2018dark} and illustrate how dark patterns can affect the users' privacy~\citep{mathur2021makes}. Remarkably, 
we extend these taxonomies, due to the mobility characteristics of AR as well as 
other essential user-oriented factors, such as safety and privacy. 


\section{Scenario Construction}~\label{sec:scenario_construction}
In this section, we utilize \textit{scenario construction} -- motivated by previous works~\citep{mhaidli2021identifying,speicher2018xd,di2020ui} to evaluate the dark pattern techniques that can be applied in AR applications. 
This approach can help us understand the possible impact of dark patterns and explore potential problems with such techniques in AR.

Scenario construction is an effective tool to analyze the impact (in our case) of dark patterns in AR technologies. Previous works have used this method to describe and evaluate the impact of future technologies~\citep{mhaidli2021identifying,betten2018constructing,reijers2018methods} and their ethical impacts on society~\citep{mhaidli2021identifying}. In our case, we construct each scenario by assigning one of the existing categories in the aforementioned dark pattern taxonomies and illustrating the manipulative mechanisms.

In our case, the proposed scenarios are used to highlight the potential privacy threats that dark patterns can arise on users in AR environments. 
We aim to construct scenarios that are easy to understand, and 
reflect techniques that are similar to the ones currently used in other environments (e.g., websites, mobile applications), so we can have more accurate examples~\citep{wright2014ethical}. Although the commonly used approaches in non-AR-based interfaces, such as highlight buttons, can also be used in many instances of the AR ecosystem, we propose more innovative techniques that combine AR applications' mixed (physical-virtual) nature. We also constructed the scenarios according to popular applications, such as games and furniture applications, to achieve a more realistic analysis of the effects of dark patterns in AR environments. We use the following prompts to develop our initial set of scenarios, where dark patterns can influence the users' decisions in AR environments:

\begin{itemize}
    \item Can we use existing dark patterns techniques in AR?
    \item How can we take advantage of AR's immersion, ubiquity, and mobility to increase the effects of dark patterns?
    \item How the dark pattern technique can influence users' decisions?
\end{itemize}

These prompts were revised and iterated over the initial set of scenarios (following the aforementioned taxonomies and features of AR). 
We condense some of the following scenarios into real prototypes to analyze the users' perceptions of dark patterns or manipulative designs in
AR interfaces. 

\subsection{AR features}
Following previous works~\citep{mhaidli2021identifying,barnes2016understanding,roesner2014security,betten2018constructing,bosch2016tales,greenberg2014dark,speicher2018xd}, we identify these defining characteristics of AR that differ from other modalities (e.g., websites): immersiveness, pervasiveness, and realism. We enumerate the above characteristics according to the current technologies, and we acknowledge that this may vary in the future. 
These traits, which come with AR environments by nature, can pose novel risks~\citep{roesner2014security} to users compared to the dark pattern taxonomies that are currently used for UIs in mobile apps~\citep{di2020ui,bosch2016tales} and websites~\citep{utz2019informed,mathur2019dark,machuletz2020multiple}. 

\bparagraph{Immersive.} In the context of AR, immersive can be seen as a seamless 
blend of the virtual world with the physical surroundings. The digital enhancements or virtual objects are smoothly rendered over the physical or real world without users' awareness. The users will believe that 
digital and virtual are part of the physical (real) scene. The level of immersion will depend on the technologies used to render the digital world and how the users interact with it. For example, most of the current mobile AR applications are displayed on the screen of a smartphone which reduces the immersive as users have a screen-through device, and the interactions are simple touches on display. Despite the current technological limitations, AR is still more immersive than non-AR mediums~\citep{lombard2001interactive}. Therefore, it is important in the context of dark patterns to include constructed scenarios and examples that use immersive AR applications.

\bparagraph{Pervasively.} The ubiquity nature of AR applications can be an important feature that is always around the users. We can showcase an example with Google Glass or similar AR smart glasses and the privacy issues that appear around them due to the continuous sensing of the user's environment~\citep{roesner2014security}. By being worn or used anywhere and anytime, AR applications open new privacy threats to users, such as collecting their surroundings (e.g., friends, and personal spaces). Although other modalities pose a privacy threat when dark pattern techniques are implemented, in the case of AR, the risks and possible harms are direr due to the ubiquity nature of AR.

\bparagraph{Realism.} We can relate realism to immersiveness as both combined can significantly improve the overall user experience in AR environments. A realistic and immersive AR application can pose a risk to 
the safety of users, as they will have issues discerning 
digital from real~\citep{jung2018ensuring}. Dark pattern techniques can involve scenarios where the users are completely unaware of their surroundings, leading them into unsafe situations (e.g., crossing the road without checking the traffic light). 
The danger that dark patterns leverage AR technologies to distort the users' reality and influence their behavior is very relevant. 
To synthesize the AR scenarios, we use the above distinctive features of AR environments 
in the following sections.


\subsection{Scenarios}

\begin{figure}[t]
  \centering
        \includegraphics[width=1\columnwidth]{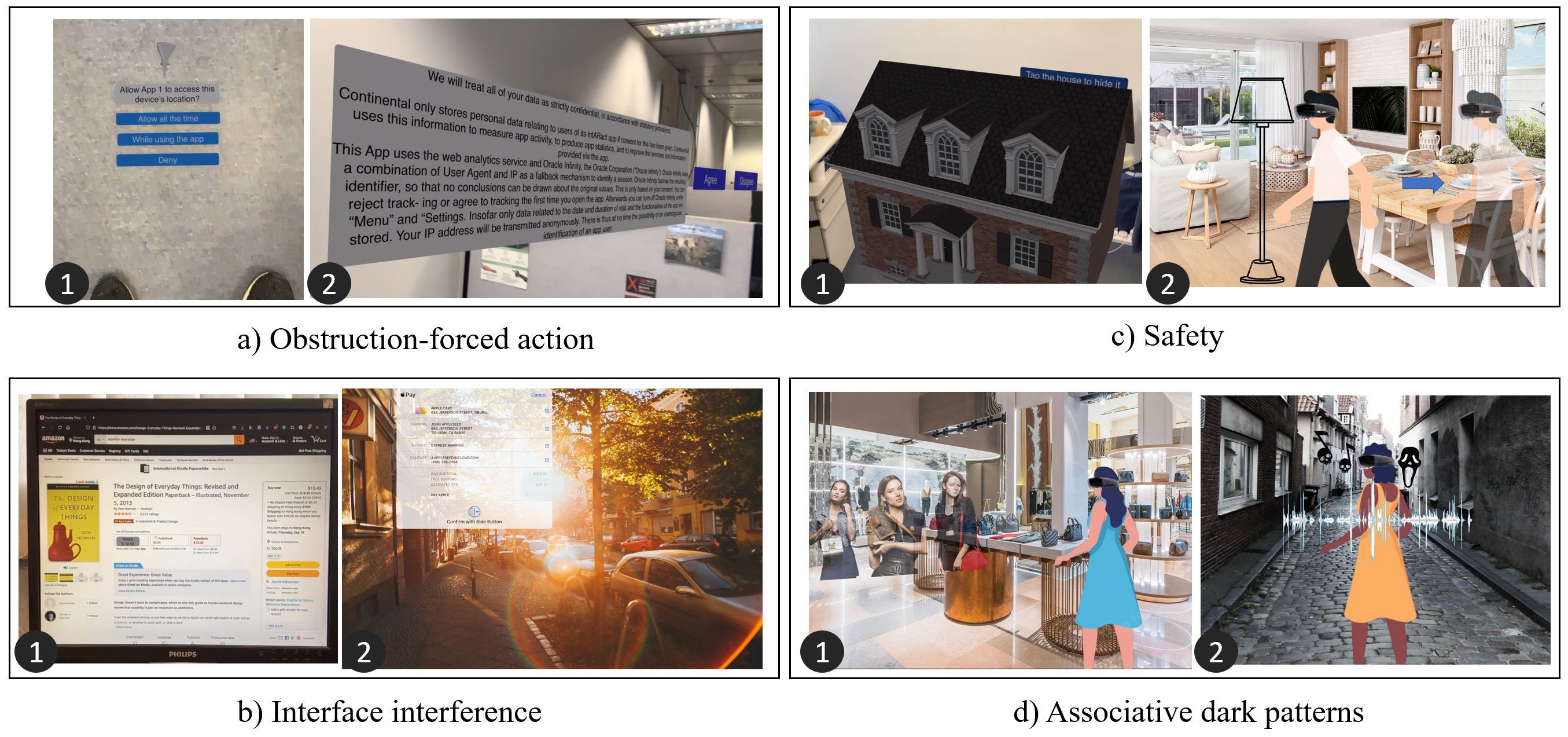}
        \caption{Examples of dark patterns in AR scenarios. a)\ding{172} The virtual interface on the ground obstructs the user's way forward; a)\ding{173} A virtual interface that appears in an inappropriate location, forcing the user to walk to a special location to unsubscribe; b)\ding{172} A virtual pricing icon blocks out the physical price tag; b)\ding{173} A blinding halo of light obscures the information on the interface; c)\ding{172} The virtual item (house) is unnecessarily bulky in the users’ view. As such, it enforces the user to step backward in order to visualize the whole item. As a result, the user may be at risk of hitting objects in the blind spot of the field of view; c)\ding{173} User crashes into a tangible table to avoid the virtual lamp; d)\ding{172} Users see virtual celebrities using luxury goods while in a luxury store; d)\ding{173} Some mood-altering (e.g., horror) sounds and vibrations (head-mounted AR controllers or some wearable devices) can influence the user choices.}
        \label{fig:scenario_construction_new}
\end{figure}

As we have described above, AR applications go beyond more traditional interfaces, and the taxonomies from an interface perspective can limit the possibilities of dark patterns in AR. We can still sample a few scenarios according to some categories of this taxonomy. Still, 
following other works, such as~\citep{greenberg2014dark,mhaidli2021identifying}, we can observe that AR `dark' designs can go beyond traditional interfaces. Figure~\ref{fig:scenario_construction_new} shows some examples of our scenario construction.





    \bparagraph{Obstruction-forced action,} these techniques make use of blocking or forcing particular actions on users (e.g., window pop-ups, mandatory registration to read a news website), drive or guide users to do some behavior.

    \begin{quotation}
    \textit{A retail store has an AR application that requests users to move to a particular physical location (e.g., a store) to unsubscribe from 
    their newsletter. The retail store uses this unsubscribing method to force users to go to their physical store, raising the possibility that the users might buy something once inside the store.} 
    \end{quotation}
    
    \cparagraph{Construction Process.} 
    Under this lens, we can observe the potential burden on users due to walking to a particular location or the change in their path due to virtual obstacles. Moreover, this technique can be combined with other mechanisms, such as sneaking or obstruction, to force users to move to a particular physical location. In contrast, the different mechanisms can enable more sophisticated dark patterns~\citep{roesner2014security}.

    \bparagraph{Interface interference,} 
    this is one of the most common techniques manifested in interfaces to deceive users when (for example) they agree to the cookie consent notice (a highlighted accept button)~\citep{bermejo2021website}. 
    
    \begin{quotation}
    \textit{An online retail store develops an AR application to allow customers to buy items by focusing on them in the physical world using the AR field of view. When the consumers want to select the delivery method, the interface uses the physical world objects to obfuscate the delivery method without extra charges. Without being noticed properly, the consumers will choose the delivery method (with extra charges) within the ones they can see.} 
    \end{quotation}
    
    \cparagraph{Construction Process.} To construct this scenario, we first consider the applications that use highlighted buttons or hidden menu options (using links instead of buttons) to force users to select the desired (by third parties) options~\citep{bermejo2021website}. 
    In the case of AR scenarios, interaction designs are no longer limited to 2D interfaces. More importantly, designers and developers can further implant dark patterns in physical objects to blend the user's surroundings. 

     \bparagraph{Safety.} The pervasive, mobile, and realistic natures of immersive scenarios in AR applications open new threats to the safety of users. 
            
        \begin{quotation}
        \textit{An AR game uses realistic virtual interfaces and objects to provide an immersive experience. This application looks 
        real and natural to users. As a result, the users 
        cannot distinguish the physical objects from the virtual ones. The application can then uses these virtual objects to create virtual obstacle near the physical ones. Once the user tries to avoid the virtual obstacle will trip over 
        the physical one in the vicinity.}
        \end{quotation}
        
        \cparagraph{Construction Process.} To construct this scenario, we follow previous works~\citep{roesner2014security,lee2021adcube}, where authors discuss the risks to user safety 
        that immersive technologies such as AR can have on users. Contrary to more traditional environments such as websites, in the case of AR scenarios, the dark pattern causes them to harm themselves by hiding real objects (e.g., a table) under virtual scenarios, or covering their field of view, and consequently forcing them to walk with an incomplete perspective of the physical world around them. 
        
    \bparagraph{Associative dark patterns.} These dark patterns can tie emotions to products in order to influence users' purchase decisions.  
    
        \begin{quotation}
        \textit{A perfume brand uses the augmented view of an AR application to display the perfume, being purchased by other people like close friends and influencers, through the lens of AR. 
        That is, the 
        application associates the person with perfume. A user might be unsure about what perfume to buy, but the application will display the above perfume-person association 
        in the buyer's surroundings. This social association, as augmented information, will influence the buyer to buy a specific perfume related to someone in the buyers' social circle or someone (bystander) that he/she likes.} 
        \end{quotation}
        
        \cparagraph{Construction Process.} To construct this scenario, we follow the associations that brands utilize to sell their products, such as campaigns that use models and particular features that trigger some emotions in buyers~\citep{kidane2016factors}. 
        For example, watches usually are associated with luxury and beauty. 
        The use of feedback, such as audio or haptic, whenever users are nearby a particular place (e.g., a retail store) can trigger emotions (e.g., fear) towards it. 
    

Moreover, the nature of AR applications and their usage in the physical world can lead users to become not only subjects of undesired data collection practices, but also threats to 
their personal safety~\citep{jung2018ensuring}. For example,  (e.g., hiding 
real objects with virtual ones so users will bump into them - so-called diminished reality~\citep{herling2012pixmix,leao2011altered}). 
Similar to their findings (though in VR) in the work~\citep{dao2021bad}, users are unaware of their surroundings or the virtual objects (original, fake copy, interactions with virtual objects)~\citep{lee2021adcube}.

\subsection{Limitations}

Although we construct our scenarios using previous taxonomies~\citep{mathur2021makes,gray2018dark}, we need to reinterpret some of the definitions, and use cases, and extend them for immersive experiences. We also restrict the prototypes to reduce the redundancy of dark pattern examples and focus on simple scenarios, so the participants in the online study can easily (if any) detect and identify these `malicious' approaches. We acknowledge that the design space of the constructed dark pattern scenarios can be broader. In this work, we aim to construct these scenarios following current dark pattern trends so the participants can identify these mechanisms.

Another limitation of our categorization is the lack of an in-depth study of current dark pattern practices in AR applications. We do not include such classification of instances of dark patterns due to the 
premature ecosystem of AR applications and the 
limited capabilities of headset devices, i.e., smartphones are the primary testbed of AR. 
Future extensions of this work will aim to categorize the current dark pattern used in AR applications accordingly. 

Using these scenarios, we developed an initial set of AR scenarios where dark pattern techniques are implemented. We avoid overly complicated AR examples to provide a subset of distinct scenarios that are easy to understand. The following section describes the user study, using VR environments as mockups of scenes being constructed, that analyzes the effects of dark patterns.

\section{User Study}~\label{sec:user_study}
We carried out a two-stage user study to analyze in detail the effects of dark patterns in AR scenarios.






\subsection{Participants and apparatus}

\bparagraph{Participants.} We recruited 
\textcolor{black}{15 (10M/5F, $\overline{age}$ = 24.27),} participants using a consecutive sample. All the participants have a high education level or higher. Seven participants reported having some XR experience, such as VR or AR games. All participants self-reported an emphasis on privacy, with a mean value of 4.53 ($SD = 0.74$, on a 5-point Likert scale). All participants reported good vision and come from Asian cultural backgrounds.

\bparagraph{Apparatus.} We built our mockup virtual scenarios using an Oculus Quest 2 and its controllers (1832 $\times$ 1920 pixels per eye, 120 Hz, 90 FOV). We developed the three scenarios using Unity\footnote{\url{https://unity.com/}} and the SteamVR plugin\footnote{\url{https://assetstore.unity.com/packages/tools/integration/steamvr-plugin-32647}}. The application was running on a stationary PC.

\subsection{Design and structure} 

 Before reporting the design study, we describe the different implemented AR applications according to the subset of scenarios described during the scenario construction (Section~\ref{sec:scenario_construction}). 

\begin{figure}[t]
  \centering
    \begin{subfigure}[b]{0.4\columnwidth}
        \includegraphics[width=\columnwidth]{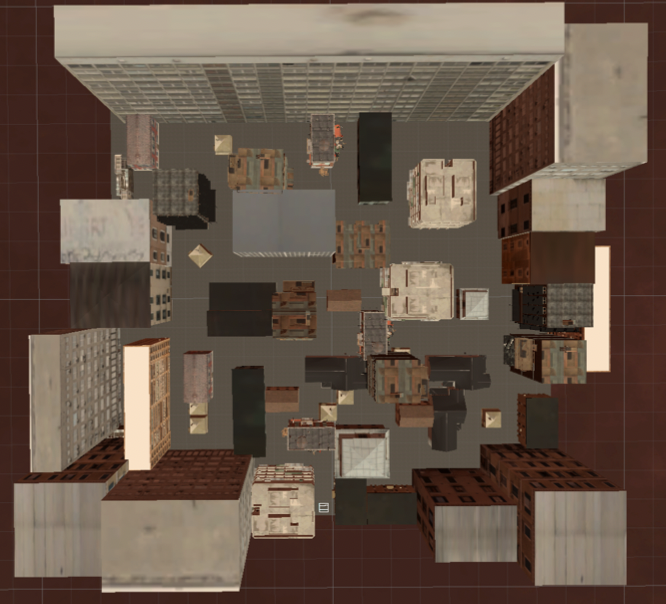}
        \caption{Aerial view.}
        \label{fig:user_study:virtual_maze:aerial}
    \end{subfigure}
    \begin{subfigure}[b]{0.4\columnwidth}
        \includegraphics[width=0.9\columnwidth]{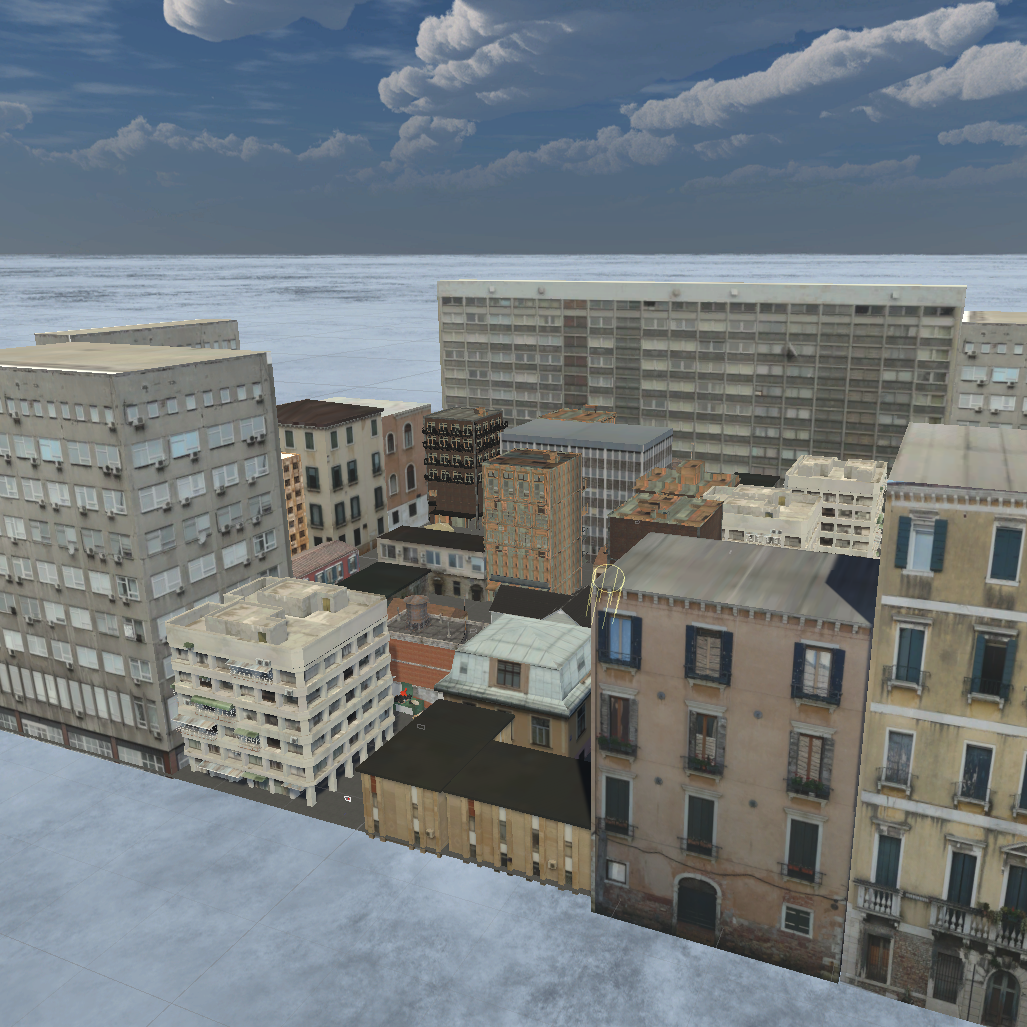}
        \caption{Side view.}
        \label{fig:user_study:virtual_maze:side}
    \end{subfigure}
    \caption{Screenshots of the virtual maze. In Figure~\ref{fig:user_study:virtual_maze:aerial}, we can see an aerial view, the notices will pop up at specific locations in the maze. 
    Figure~\ref{fig:user_study:virtual_maze:side} depicts the side view of the virtual maze.}
    \label{fig:user_study:virtual_maze}
\end{figure}

\begin{figure}[t]
  \centering
    \begin{subfigure}[b]{0.45\columnwidth}
        \includegraphics[width=0.92\columnwidth]{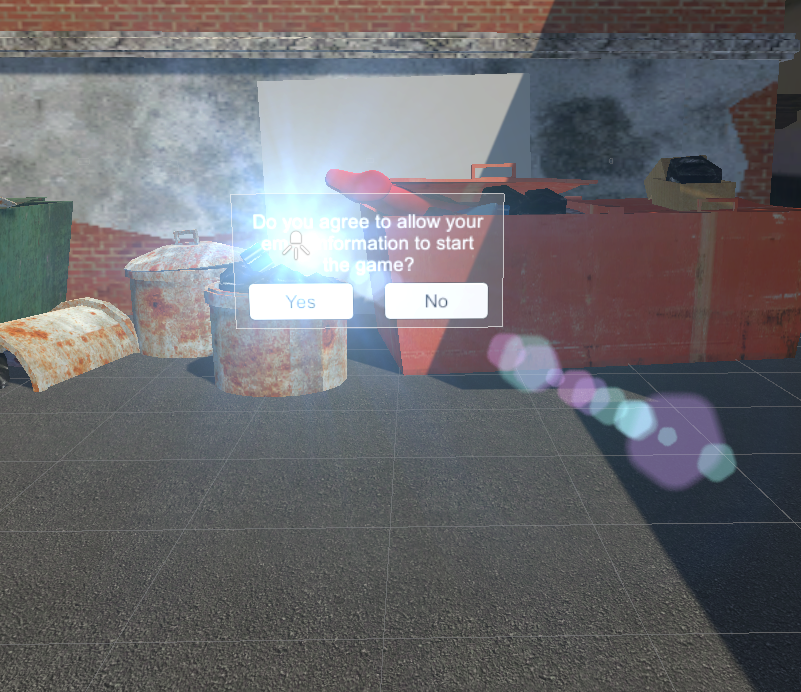}
        \caption{Lighting inter.}
        \label{fig:user_study:conditions:light}
    \end{subfigure}
    \begin{subfigure}[b]{0.4\columnwidth}
        \includegraphics[width=0.94\columnwidth]{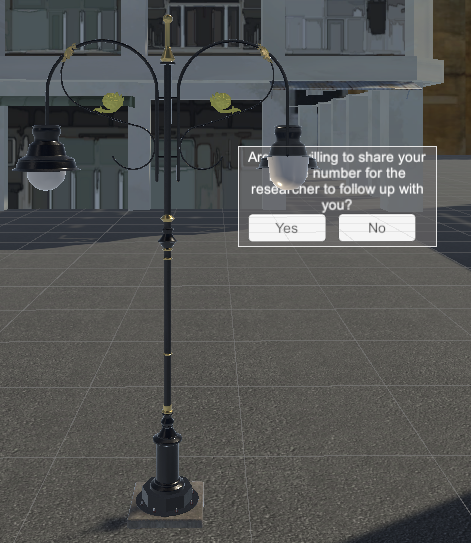}
        \caption{Object inter.}
        \label{fig:user_study:conditions:object}
    \end{subfigure}
    \begin{subfigure}[b]{0.5\columnwidth}
        \includegraphics[width=0.9\columnwidth]{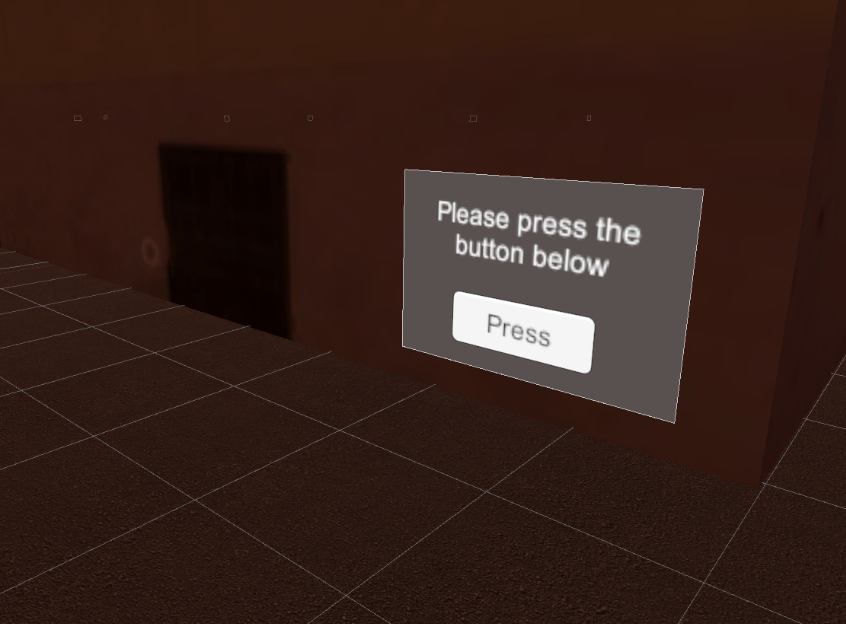}
        \caption{Haptic nagging.}
        \label{fig:user_study:conditions:haptic}
    \end{subfigure}
    \caption{Screenshots of the different conditions that participants will see during the experiment. We can observe the obfuscated words in Figures~\ref{fig:user_study:conditions:light} and \ref{fig:user_study:conditions:object}. Figure~\ref{fig:user_study:conditions:haptic} illustrates the haptic condition, where the notification is placed on one side of the street.}
    \label{fig:user_study:conditions}
\end{figure}

\bparagraph{Conditions.} Three scenarios were designed to explore dark patterns in AR. These dark pattern design 
approaches are a sampling of what the future of dark patterns in AR could be. These 
approaches are still grounded by the current technologies in AR environments. We use VR as a tool to immerse participants in the respective scenarios, as these VR mockups can emulate the AR environment in the virtual scene (see Figure~\ref{fig:user_study:virtual_maze}).


\begin{itemize}

    \item \textbf{Lighting interference.} In this scenario, we follow similar design approaches: 
    interface interference, sneaking, and obstruction, as proposed in previous works~\citep{gray2018dark,darkPatternsBrignull}, where UI elements (e.g., buttons with different colors) influence the users' decisions. In this case, the message will hide specific words in the questions using strong lighting conditions (e.g., halo), see Figure~\ref{fig:user_study:conditions:light}. The participants cannot see clearly the whole question and they are forced to move around in order to visualize the complete notice.
    
    \item \textbf{Object interference.} Similar to the previous interference, in this case, we use physical objects (street lamps in our VR mockup) to hide specific words in the messages or notifications (Figure~\ref{fig:user_study:conditions:object}). 
    We showcase two possible approaches that AR developers can use to manipulate users' decisions with a combination of virtual and physical objects.
    
    \begin{figure}[t]
     \centering
        \includegraphics[width=1\columnwidth]{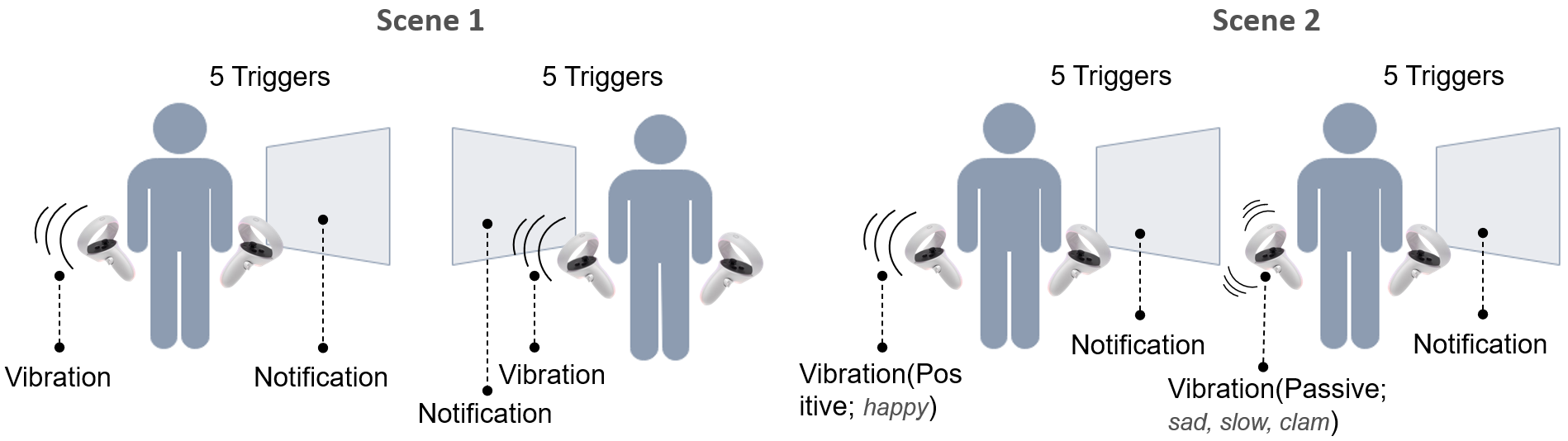}
        \caption{\textcolor{black}{Illustration setup for 
        haptic grabbing scenes. Each participant experiences triggers and notification interfaces for aligned/opposite directions in each scene.}}
        \label{fig:user_study:Haptic_Exp_Design}
    \end{figure}
    
    \item \textbf{Haptic grabbing.} This condition indicates the impacts of haptic feedback, as rendered by VR controllers, 
    to grab the participant's attention to the right or left of the virtual map, producing corresponding stimuli on the left- or right-hand controller, respectively (Figure~\ref{fig:user_study:conditions:haptic}). We designed two scenarios, see Figure~\ref{fig:user_study:Haptic_Exp_Design}: \textit{scene 1}, we evaluate the effects of haptics on participants' movements; \textit{scene 2}, we study the effect of positive/negative haptic stimuli to grab participants' attention towards the notice. 
    Attention grabbing can interfere not only with the decision-making process of users but -- in this case -- reduce the probability of viewing 
    the notices on the virtual map. 
    Motivated by the proposed scenario in~\citep{greenberg2014dark}, we evaluate the efficacy of haptic feedback to control users' movements in our VR mockup. For example, this could allow new manipulative designs that leverage 
    haptic notifications to prevent 
    users from looking at a nearby physical store. 
    In our study, the participants were told that they would receive haptic feedback at particular locations in the virtual maze and with different controllers (left or right).
    

\end{itemize}

According to prior studies~\citep{acquisti2005privacy}, privacy is an important concern of users and one of the main goals of 
dark pattern designs~\citep{mathur2021makes,bermejo2021website,di2020ui,gray2018dark}. For both lighting interference and object interference conditions, we include different questions regarding the sharing of personal information of the participants (e.g., online social accounts). With these questions, we aim to outline a more realistic scenario and, thus, users' behavior~\citep{distler2021systematic}.

\bparagraph{Questions (object and lighting interference conditions).} We designed the following questions according to the privacy risks that the participants would face in a university environment. We show the participants a website with a leaderboard that includes information about the previous participants, such as their GPA and email. We aim to provide a more realistic environment using these questions and the website. Hence, the participants believe that their results (time) and personal information, such as their GPA and name, will be publicly accessible. The participants can answer yes or no to the following questions:
\label{sec:questions}
\begin{itemize}
    \item Are you willing to share your \textbf{Student ID} with other players at the end of the game?
    \item Are you willing to share your \textbf{GPA} on our web?
    \item Are you willing to share your \textbf{social accounts} on our web?
    \item Would you like to share your \textbf{game ranking} on our web?
    \item Do you agree to allow your \textbf{email} information to start the game?
    \item Are you willing to share your \textbf{phone number} for the researcher to follow up with you?
\end{itemize}

Where the bold keywords are obscured by strong light, all collection processes are fake, but participants are not informed in advance. For all questions, the user only has to select yes or no in the interface.

\bparagraph{Haptic grabbing condition.} The scenario will pop up (at specific locations in the maze) a notification with the following text: \textit{Please press the button below} to confirm the participants have seen the notification, see Figure~\ref{fig:user_study:conditions:haptic}.


\bparagraph{Pilot Study.} We design a pilot study to evaluate the valence and arousal (VA) of the controllers' haptic feedback for VR. The pilot study aims to evaluate the effect on participants' emotions (valence and arousal), driven by the haptic stimuli (i.e., using vibroactuators from the Oculus Quest 2 controller). We design six different haptic feedback following the dark patterns in~\citep{makela2021hidden,yoo2015emotional}. The parameters used to render the vibrotactile feedback are twofold~\citep{rovers2004him}: (I) intensity, three different levels of intensity relative to the default provided by the controller (1) (low (0.5), medium (0.7), high (1)); and (II) frequency, with three different types relative to the default frequency (1) (continuous (1), slow (1/2), and fast (1/4)). We evaluate the responses of nine participants from our university campus (4M/5F, $\overline{age}$ = 21). The participant can experience the same haptic pattern multiple times until they felt that the haptic feedback was completely perceived, at which moment the participant pressed the trigger button on the controller to pop up the Self-Assessment Manikin (SAM) screen~\citep{bradley1994measuring}, which helped the participant determine the emotional feedback at the moment (VA). We select the two most distinctive (according to VA) haptic feedbacks (\textit{happy}, \textit{sad}) for our `haptic grabbing' condition. We aim to study whether an effect on users' behavior exists or not, according to the VA assigned for each of the two selected haptic feedback.

\subsection{
User experiments}

\bparagraph{Task.} Participants were asked 
to solve the maze (find the exit) in the shortest time possible. Inside the maze, they had to answer several questions that would appear at particular locations. For the lighting and object interference scenario, the participants responded to 6 questions in Section \ref{sec:questions} in particular locations of the virtual maze. The areas where notifications were triggered are all on the path to the correct maze exit. For the haptic grabbing scenario, participants were requested to press the button in the notification interface they found. The task was accomplished 
when the participants found the exit of the maze or spent more than 
\textcolor{black}{5} minutes in the maze.

\bparagraph{Procedures.} Participants provided informed consent to participate in this study. Then, we introduced to the participants task and showed them a website that included 
a list of the fastest (fake participants but the participants were not informed.) times to solve the maze. We counterbalanced the order of the three different scenarios. At the end of the experiment, we asked the participants whether they noticed any possible malicious design in the scenarios and asked them to describe them (if any). Finally, the participants responded to demographic-related questions.

The participants took between 16 and 19 minutes to complete the whole experiment. They were rewarded with snacks, sweets, and soft drinks after completion.


\subsection{Results}\label{ssec:final-test}

\bparagraph{Lighting and object interference.} We observed that 71.1\% of the responses of the participants were mainly yes when we asked them about sharing any personal information in these virtual scenarios. Only one participant (P12) responded no more times (66.67\%, n=4) than yes (33.33\%, n=2) to the displayed questions. These results show the possible impact of the dark pattern (e.g., lighting interference) on the participants' responses. 

\begin{figure}[t]
  \centering
        \includegraphics[width=0.6\columnwidth]{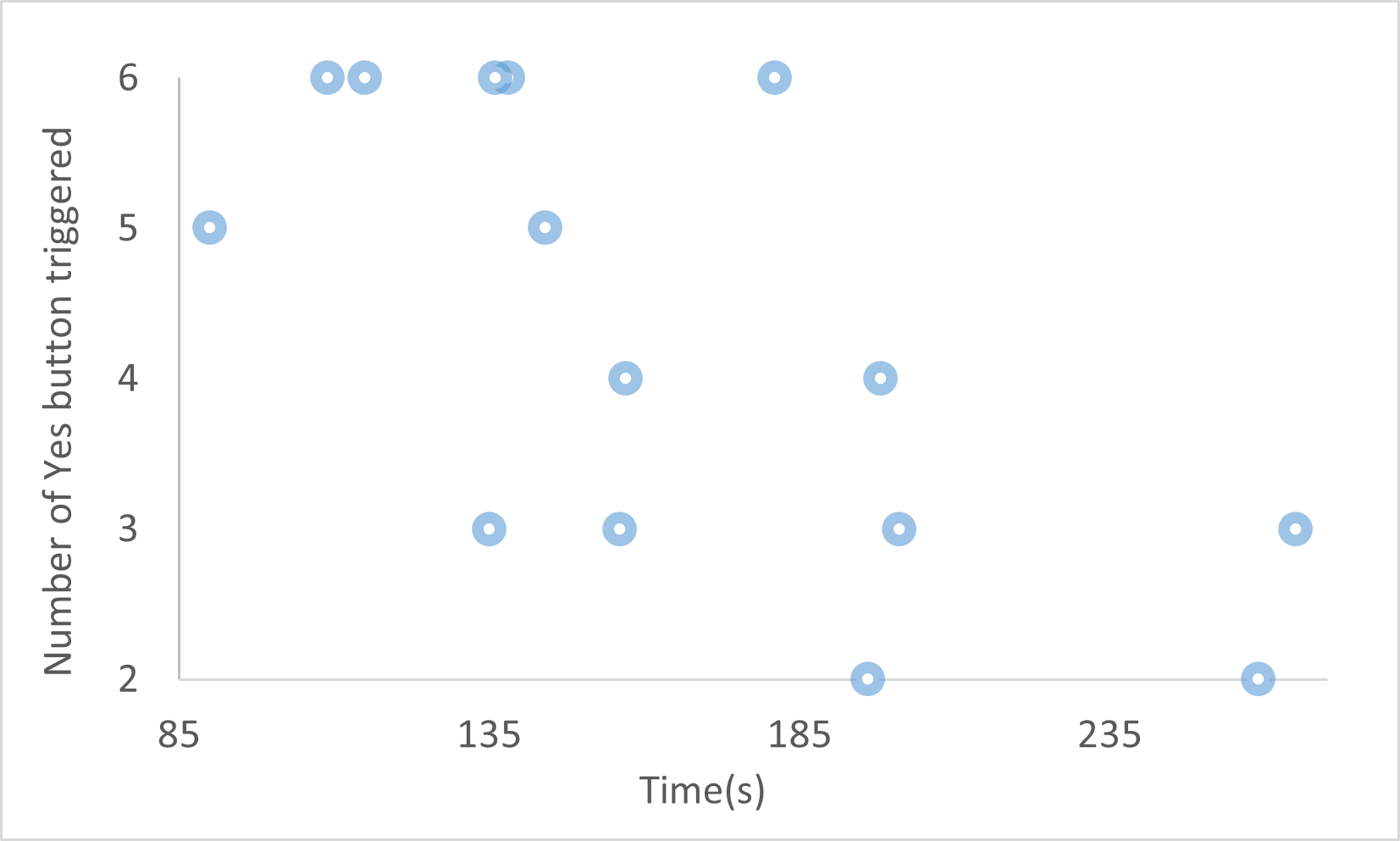}
        \caption{Lighting and object interference results: \textcolor{black}{The time it took participants to find the correct solution to the maze was negatively correlated with the number of times they encountered privacy questions in the maze and triggered the "Yes" button. With
        glare and objects in the maze, participants who wanted to avoid more privacy risks 
        consuming 
        their time going through the maze.}}
      
        \label{fig:user_study:Light_Result}
\end{figure}

The time participants spend solving the maze and the results of answering the questions in the scene are recorded. Spearman’s rank correlation was computed to assess the relationship between the number of $Yes$ buttons triggered and the time to solve the maze. There was a significant negative correlation between the two variables ($r(13) = -.63, p = .006$). The results are shown in Figure~\ref{fig:user_study:Light_Result}.

\textcolor{black}{12 out of the 15 participants were affected by the bright light on the interactive interface while solving the maze, and 7 of them reported directly to the experimenter that the light made them difficult to read the text. 
Also, the remaining 5 had the behavior of moving around the virtual environment in an attempt to avoid the glare. The dark patterns (object and lighting interference) increase the time that each participant spends answering the questions. When we asked the participants at the end of the experiment whether there was any malicious design in the scenario, 8 participants reported that there was a malicious design. Moreover, 7 of these 8 participants reported the glare/light as the malicious design used in the scenario.
Two participants mentioned that the street light in the scene was strange and seemed to have been placed there intentionally, and one of them thought that the street light also obscured the interface.} The participants reported the use of some manipulation techniques when explicitly asked. However, the results show that the participants fell for the malicious design (responding yes to the displayed question).

\bparagraph{Haptic grabbing.} Participants were required to experience two scenes under this condition. 
In each scene, a participant encountered 10 triggered notification (as vibration) with corresponding visual pop-up information (i.e., notification interfaces). 
Regarding the first scene, 
5 out of 10 notification interfaces deliver the vibration on the same side with the visuals being popped up. The remaining half of notification interfaces make opposite direction, i.e., the right-hand-side controller vibrates but notification visuals appear on left-hand side, or vice versa. We note that the participants were informed about the notification interfaces. They had to press the button on the controller to confirm that they spotted the visual notifications. 
Accordingly, the second scene aims to examine the vibration type of positive and negative modes, under the condition of that all ten notification interfaces maintain an opposite direction between the vibration and notification visual. Five of the notification interfaces refer to the positive vibration and the remaining five notification interfaces represent the negative counterpart. Figure~\ref{fig:user_study:Haptic_Exp_Design} shows the details of the scenes setup.  
In the two above scenes, the number of notification interfaces pressed by the participants was recorded.

\begin{figure}[t]
  \centering
        \includegraphics[width=0.6\columnwidth]{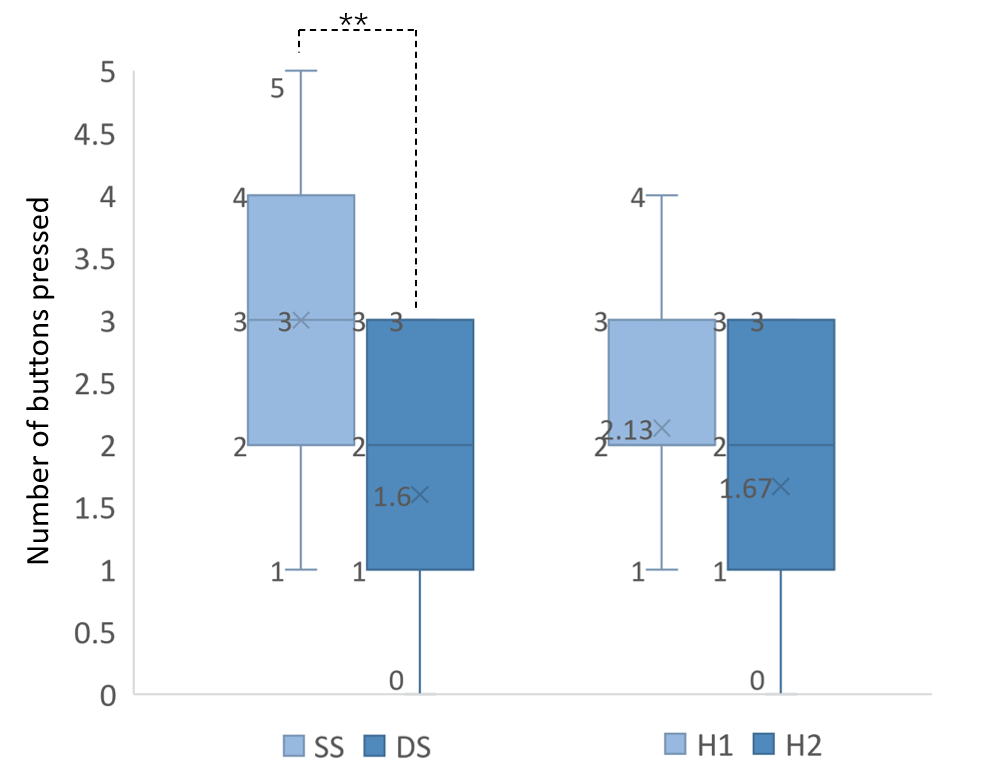}
        \caption{Haptic grabbing result, (p $<$ .05 ($\ast$), p $<$ .01 ($\ast\ast$)), where \textit{SS} corresponds to rendering the haptic on the same side as the notification and \textit{DS} to different side. \textit{H1} corresponds to the B3 (happy active), and \textit{H2} to haptic B5 (sad, slow, calm).}

        \label{fig:user_study:Haptic_Result}
\end{figure}

\textcolor{black}{
A Shapiro-Wilk test showed that the data satisfied a normal distribution. 
A one-way ANOVA showed a significant difference in the number of notification interfaces found by participants in the different conditions ($F(3,56) = 5.25, p < .005$). Tukey’s HSD Test revealed that the number of notification interfaces found by participants was significantly higher when the vibration and notification locations were on the same side (SS) than on the different side (DS) condition ($p = 0.005$). In contrast, different vibration haptic patterns did not have a significant effect on the number of notification interfaces that participants were aware of ($p = 0.558$). The results are shown in Figure~\ref{fig:user_study:Haptic_Result}.}

\textcolor{black}{Except for two participants with very serious cybersickness, the remaining 13 participants exhibited the behavior of stopping to look around/seek when the controller vibrated. The results of the questionnaire indicated that 5 participants believed there were malicious designs in the scenario. 
Remarkably, 4 of them pointed to vibration as a possible malicious design. One of the participants noted that \textit{"I would unconsciously move in the direction of the controller vibration"}, and another said that \textit{"The controller vibration seemed to manipulate me intentionally"}.}

\bparagraph{Post-experiment interview.} We briefly interviewed participants in the questionnaire who believed that light and vibration were malicious designs in the scenario. First, all participants who believed that light was a malicious design agreed that the use of bright light to obscure critical information was an invasion of their privacy, while only one of the participants who believed that vibration was a malicious design considered it to be an invasion of their privacy. In addition, all participants agreed that the vibration design violated their rights. 
A participant said, \textit{``The vibration design makes me miss information that I am entitled to see''}. All participants agreed that this kind of malicious interaction design is likely to appear in future AR interaction designs.




\section{Discussion and Research Challenges}~\label{sec:discussion}
The results of our scenario construction and user study show the impact of dark patterns in AR scenarios. Visual-based manipulations (lighting and object interference) affect the participants' responses. Moreover, there is an impact on the participants' time to solve the maze when using these dark pattern designs, as the participants that did not agree with sharing their personal information (try to read the whole message) spent more time solving the maze. In the case of haptic grabbing, we saw an impact of the haptic stimuli on participants' movements. The haptic feedback nudges the participants toward the direction of the controller rendering the stimuli.

\subsection{Dark Patterns in immersive experiences}

Novel 
and more effective dark patterns can be implemented in AR due to the immersive and realism of the environments. These improved user experiences can enable more sophisticated approaches to alter users' behaviors. In comparison with other traditional user interfaces (e.g., desktops, phones), dark patterns in AR can have a greater impact on the users' decision process~\citep{mhaidli2021identifying}. Most of these traditional approaches meticulously design user interfaces to `trick' and influence users without their awareness~\citep{gray2018dark}. In the case of AR applications and hyper-realistic virtual objects, the users will not be able to distinguish them from physical ones~\citep{ruth2019secure}. \textit{How 
can we tell users that a hyper-realistic virtual object is not genuine without hindering the user experience and immersion?}

Besides the realism of virtual objects, the immersiveness and interactivity of these environments can also impact the effectiveness of dark patterns. 
Similar to the effects of audio on the shoppers' buying intentions in retail stores~\citep{milliman1982using}, AR applications can create immersive dark patterns using visual, audio, and haptic feedback. The AR ecosystem can augment the effect of dark patterns into influencing users' behaviors. It is necessary to 
analyze the main factors influencing users when dark patterns (using visual, audio, and haptic) are applied in AR environments, and how can we protect users in these immersive experiences.

Finally, AR applications allow users to interact with their interfaces using more natural approaches (natural interactions)~\citep{shatilov2019emerging}. These natural interactions can also be the subject of dark pattern techniques. 
For example, when a user clicks a UI button via gaze or using the index fingertip as a controller. In this case, a dark pattern can hide (hijack) the authentic cursor behind a virtual object and use a fake cursor to click another UI interface (e.g., an ad banner)~\citep{lee2021adcube}. These virtual elements, being inserted during the input interactions of users, require framework-driven regulations 
within the AR ecosystem. Following the work in~\citep{lee2021adcube}, future AR platforms should consider the defense against dark pattern attacks, whose techniques are implemented during the interaction process (e.g., cursor movement). 

Moreover, we can see dark patterns in AR as interface-dependent, where the different screens/field of view options can have different implications for the implementation of dark patterns. This, together with the multi-modal nature of AR applications (visual, auditory, and haptic), can open new approaches for the design of dark patterns beyond traditional systems such as smartphones and computers.


With this work, we aim to bridge the gap between current technologies for AR and the possibilities of dark patterns in AR described in the constructed scenarios. We describe the future challenges and approaches that such manipulative scenarios can pose in the development of AR applications. In the following paragraphs, we highlight the main research challenges and intervention mechanisms that could be implemented to counterbalance the use of dark patterns.

\subsection{Privacy of AR}

The sensing requirements of AR applications, such as capturing camera frames, open privacy risks that differ significantly from traditional interfaces (i.e., desktops, phones). As mentioned in previous works~\citep{guzman2021unravelling,de2019security,roesner2014security}, the scanning capabilities of devices used in AR ecosystems raise several privacy concerns and challenges to protecting users' privacy. For example, the spatial information captured from these devices (point clouds) can leak information about the personal space of a user (e.g., bedroom size) and objects (e.g., furniture) belonging to a particular user~\citep{guzman2021unravelling}. Designers can develop dark pattern techniques that force users into capturing more information about their surroundings without their awareness. Obviously, some of these designs are only intended to drive/change specific user behaviors and are not malicious or intended to manipulate users. However, they can still raise privacy concerns among users.

\subsection{Safety of AR}

Malicious designs can cause critical risks to the safety of users in AR environments~\citep{lebeck2017securing, roesner2014security}. 
For instance, AR cues become the obfuscation visually to a particular street in the physical world. When a user is crossing a road, the obfuscation can put the users in danger due to reduced awareness of incoming vehicles. 
In comparison with other traditional interfaces, dark patterns can become a more prominent threat for users of AR applications. 

The management of content and visual output of virtual objects in AR applications can guarantee 
a safer environment for users~\citep{ruth2019secure, lebeck2017securing}. 
These security and safety risks should be mitigated with platforms and regulations that control how the visual output is displayed to users \citep{james-lam} in a harm-free manner. 

\subsection{Intervention mechanisms}

In this section, we discuss the results in the scope of the problems that dark pattern arises from previous works in more traditional environments (desktop, smartphone)~\citep{bongard2021definitely,di2020ui,bermejo2021website,utz2019informed,mathur2019dark}. 
We connect the previous works regarding the issues with AR~\citep{roesner2014security,lebeck2017securing,ruth2019secure,de2019security,langheinrich2001privacy}, to 
our findings in the realm of dark patterns and AR.  
Intervention mechanisms aim to provide individuals with the tools and protection techniques they need to protect their personal information from malicious designs such as dark patterns. These intervention techniques can be developed for either the user or the environment. For example, regulators will enforce and intervene in scenarios where dark patterns can put users with AR applications at risk. 

\bparagraph{Increasing awareness.} Our results show that users are not fully aware of the manipulation techniques used in dark pattern designs in AR, following previous work relevant to traditional interfaces~\citep{di2020ui,utz2019informed,bongard2021definitely}. 
Moreover, even when users are aware of design `tricks', they do not fully understand the manipulation or its purpose (e.g., what type of information is being collected). 
In the case of AR applications, individuals' awareness and concerns regarding dark patterns are even lower due to the lack of experience with AR applications and the novelty of this environment (as we discussed in this paper), i.e., the users' technology literacy to the evolving malicious techniques. 
The addition of feedback cues, such as audio and haptic, to raise users' concerns (e.g., warnings~\citep{gray2018dark}) when a particular suspicious design is being used, can increase user awareness and their understanding of dark patterns. 
However, warning techniques can reduce their effectiveness with time (users get habituated to them)~\citep{akhawe2013alice}. Mechanisms that increase users' awareness of dark patterns should change with time (e.g., different feedback modes, not always active) so as to keep their effectiveness.

\bparagraph{Education and training.} Our results show that participants with some experience with AR applications had higher chances to detect dark patterns in our study. Moreover, following the above statement about user awareness and concerns regarding dark patterns, we believe that sustainable mechanisms to further educate users about the threats of dark patterns and how these can be detected can increase the individuals' self-protection abilities. For example, the combination of the above warning techniques, together with better education and training about the threats of dark patterns (e.g., privacy and attention-grabbing mechanisms), can allow users of AR applications to become more aware of these malicious designs.

\bparagraph{Counteracting dark patterns in the digital space.} As the results of our study show, dark pattern techniques can have a significant impact on users' behaviors. For example, attention-grabbing techniques using haptic feedback change the participants' movements in the maze, and interference approaches using lighting conditions and objects affect their responses. Moreover, these manipulative designs will overcome the countermeasures of several frameworks that protect users from malicious AR output (e.g., notifications that cover users' field-of-view)~\citep{lebeck2017securing,roesner2014security}. These frameworks should also include more control over the interface design and output of virtual elements and allow users to report and customize the AR output to limit manipulative designs such as the ones studied in this paper. Also, these countermeasures should consider the interface used to interact with AR, as we can see an interface-dark pattern correlation in the design of manipulative techniques.



As we described in our scenario construction section, the possibilities for designing AR interfaces are humongous compared to more traditional interfaces (desktops, smartphones). Even when the dark patterns are somewhat analogous to well-recognised designs from these traditional interfaces, individuals are not fully aware of them. Moreover, AR applications not only pose privacy threats to users~\citep{de2019security} but also endanger their safety~\citep{roesner2014security} when malicious techniques are applied with such a purpose (e.g., hiding real objects from users' view~\citep{leao2011altered}). Therefore, regulations, economic incentives, and ethical designs should play a mutually important role in counteracting and controlling the dark patterns in AR applications.

\subsection{Limitations of our study}

In terms of limitations, using the same physical environment can result in carry-over effects, which can impact the effects of dark patterns. 
Nevertheless, our study offers a realistic environment to study the effects of dark patterns in AR. One outcome we could not assess with the current study is the long-term effect of such dark patterns on users' everyday behavior. Our participants were from our university campus and they have some experience with AR or XR applications. Meanwhile, because our participant group was comprised of students, they would be especially concerned about the questions posed in our scenario, and all participants demonstrated a high regard for privacy issues. Thus, the participants were hesitant to reveal them; however, different groups may have varying levels of concern/non-concern/awareness/prudence regarding perceived privacy violations. Therefore, our results should be considered in the context of the participants we recruited.

We use the haptic vibroactuator device embedded in the Oculus controller for our experiment. One of the reasons we chose this option was to account for the fact that many of the most common AR headsets currently in the industry use hand controllers with vibration feedback for interactions (refer to Section~\ref{sec:devices}). We acknowledge that further studies in the haptic feedback direction should consider other stimuli areas and different haptic devices, for example, haptic gloves~\citep{nam2008smart,koyanagi2005development}, weight simulation~\citep{wang2022vibroweight} and thermal feedback~\citep{maeda2019thermodule}.

Due to the considerations mentioned in Section~\ref{sec:reasons}, we use VR mock-up to simulate AR. The emulated environment will inevitably differ slightly from the actual AR situation; for instance, the fidelity of the real environment will be higher than the street of the VR simulation, and the real world with digital overlays will provide more immersive environments to the user. If conditions permit, future researchers can construct a photo studio, e.g., a dark pattern sandbox, to conduct AR experiments for the sake of investigating user behavior in the context of dark patterns and AR, ensuring user safety and controlling environmental variables.

The current user study presents a limited type of dark pattern that can potentially appear in AR environments when they are widely adopted. The future version of this work will include more experiments and AR scenarios to provide more in-depth technical details in this direction.

\subsection{Future work}

In this work, we present novel dark pattern designs for AR according to current technologies. As revealed by 
our results, the use of alternative modalities, such as haptic feedback, can open new possibilities for dark pattern designs beyond traditional UIs~\citep{greenberg2014dark}. Similarly, audio feedback and more sophisticated haptic approaches could be used in future designs of dark patterns for AR. 
We do not include more sophisticated approaches in this work as the current technologies for haptic rendition are still in their infancy. Meanwhile, commercial approaches can be mainly found as 
vibroactuators devices, known as 
gamepads, smartphones, and smartwatches~\citep{bermejo2021survey}. The possibilities with AR applications are broader than the traditional counterparts, 
such as mobile applications and websites, where non-AR interfaces limit the interactions and techniques that dark patterns can use.

Future extensions of this work will include studying diversified devices (e.g., Microsoft HoloLens, Google Glass) available in the wide spectrum of the AR ecosystem (i.e.,, low-end to high-end devices). Moreover, we will 
extend our current work with the study of multi-modalities for dark patterns, such as haptic, audio, and visual, which can provide new perspectives on the design of dark patterns for AR, following futuristic scenarios such as the ones enumerated in~\citep{mhaidli2021identifying,greenberg2014dark,tseng2022dark}.


Finally, VR technologies and technologies in the extended reality (XR) have been getting traction recently with the announcement of the Facebook introduction of `meta'\footnote{\url{https://about.fb.com/news/2021/10/facebook-company-is-now-meta/}}. The metaverse will open new challenges to preserve users' privacy~\citep{lee2021internet}. In addition to that, ChatGPT\footnote{\url{https://openai.com/blog/chatgpt}}, an artificial intelligence chatbot built on top of the GPT-3 language model and introduced by OpenAI in November 2022, may lower the barrier to creating malicious manipulation in the future. In future applications, such chatbots are likely to be integrated with XR creation, and through such chatbots, everyone will be capable of creating AR instances more easily and conveniently in the future. These are the emerging research topics that future researchers should investigate. Future work will also include an extensive analysis of dark patterns in these virtual environments to avoid threats such as abusive ad service providers~\citep{lee2021adcube}.

\section{Conclusion}~\label{sec:conclusion}
In this paper, we present the first comprehensive overview of possible dark pattern designs for AR ecosystems and explore the potential of such technologies in AR environments. We presented and constructed several AR scenarios to explore how dark patterns could be implemented in the AR environment. Remarkably, we construct our scenarios using two taxonomies: the interface and normative perspective of dark patterns. Additionally, we conducted a two-step user evaluation. The purpose of the user study was to determine how users perceive the constructed dark patterns scenarios in the AR application. To do so, we crafted three scenarios based on possible situations in our proposed scene construction method. The results indicate that dark patterns influence user behavior, with strong lights and objects obscuring the interface compelling users to move and prolonging gameplay. The interference of vibrations can make users miss notifications. 
Our findings shed light on the users' threats from dark patterns in emerging immersive worlds, known as the Metaverse. 

\section*{Acknowledgement}

This research was partially supported by the MetaUST project from HKUST(GZ) and the FIT project (Grant No. 325570) from the Academy of Finland.

\bibliographystyle{apacite}
\bibliography{mybib.bib}

\section*{About the Authors}

\begin{description}
\item[Xian Wang] received BSc in Electronic Science and Technology from Xi'an Jiaotong-Liverpool University in 2021 and is a member of the Human-Computer Interaction Lab X-CHI. She is currently conducting her Ph.D. in Artificial Intelligence at SyMLab, Hong Kong University of Science and Technology. Her research interests include human-computer interaction, haptic feedback, virtual reality, and serious games.
\item[Lik-Hang Lee] received the Ph.D. degree from SyMLab at Hong Kong University of Science and Technology, and the BEng(Hons) and M.Phil. degrees from the University of Hong Kong. After his graduation, he was a postdoctoral researcher at the University of Oulu, Finland (2019 - 2020), and an assistant professor (tenure-track) at KAIST, South Korea (2021 - 2023). Currently, He is an assistant professor at The Hong Kong Polytechnic University, HKSAR. He has built and designed various human-centric computing systems specialising in augmented and virtual realities (AR/VR).
\item[Carlos Bermejo Fernandez] received his Ph.D. from the Hong Kong University of Science and Technology (HKUST). His research interests include human-computer interaction, privacy, and augmented reality. He is currently a postdoc researcher at the SyMLab in the Department of Computer Science at HKUST. 
\item[Pan Hui] received his PhD from the Computer Laboratory at the University of Cambridge, and has since held several prestigious academic positions. He is currently a Chair Professor of Computational Media and Arts and Director of the Centre for Metaverse and Computational Creativity at the Hong Kong University of Science and Technology (Guangzhou), as well as a Chair Professor of Emerging Interdisciplinary Areas at the Hong Kong University of Science and Technology. Additionally, he holds the Nokia Chair in Data Science at the University of Helsinki, Finland. His extensive research contributions are widely recognized, with over 450 research papers, 32 patents, and over 25,000 citations.
\end{description}

\end{document}